\documentclass[12pt]{article}

\usepackage{epsfig}
\usepackage{rotating}

\begin{document}

\title{\Large \bf Evidence for a 'Narrow' Roper Resonance -\\ The Breathing
  Mode of the Nucleon}
  
\author{ \Large H.~Clement$^1$, T.~Skorodko$^1$, M.~Bashkanov$^1$,
  D.~Bogoslawsky$^2$,\\ 
\Large  H.~Cal\'en$^3$,
F.~Cappellaro$^4$,
L.~Demiroers$^5$, E. Doroshkevich$^1$,\\
\Large C.~Ekstr\"om$^3$,
K.~Fransson$^3$, 
L.~Gustafsson$^4$,
B.~H\"oistad$^4$, \\
\Large G.~Ivanov$^2$,
\Large M.~Jacewicz$^4$,
E.~Jiganov$^2$,
T.~Johansson$^4$, \\
\Large M.~Kaskulov$^1$,
O.~Khakimova$^1$,
\Large S.~Keleta$^4$,
I.~Koch$^4$, \\
\Large F.~Kren$^1$,
\Large S.~Kullander$^4$,
A.~Kup\'s\'c$^3$,
\Large A.~Kuznetsov$^2$,\\
\Large P.~Marciniewski$^3$, B.~Martemyanov$^{11}$,
\Large R.~Meier$^1$,
\Large B.~Morosov$^2$,\\
\Large W.~Oelert$^8$,
\Large C.~Pauly$^5$, H.~Pettersson$^4$,
\Large Y.~Petukhov$^2$,\\
\Large A.~Povtorejko$^2$,
\Large R.J.M.Y.~Ruber$^3$, K.~Sch\"onning$^4$,
\Large W.~Scobel$^5$,\\
\Large B.~Shwartz$^{9}$,
\Large V.~Sopov$^{11}$,
\Large J.~Stepaniak$^7$,
P.~Th\"orngren-\\
\Large Engblom$^4$,
\Large V.~Tikhomirov$^2$,
\Large A.~Turowiecki$^{10}$,
\Large G.J.~Wagner$^1$,\\
\Large M.~Wolke$^4$,
\Large A.~Yamamoto$^6$,
J.~Zabierowski$^{7}$,
J.~Z{\l}omanczuk$^4$
 \bigskip \\
{\it $^1$~Physikalisches Institut der Universit\"at T\"ubingen, D-72076
  T\"ubingen, Germany} \\
{\it $^2$~Joint Institute for Nuclear Research, Dubna, Russia} \\
{\it $^3$~The Svedberg Laboratory, Uppsala, Sweden} \\
{\it $^4$~Uppsala University, Uppsala,Sweden} \\
{\it $^5$~Hamburg University, Hamburg, Germany} \\
{\it $^6$~High Energy Accelerator Research Organization, Tsukuba, Japan} \\
{\it $^7$~Soltan Institute of Nuclear Studies, Warsaw and Lodz, Poland} \\
{\it $^8$~Forschungszentrum J\"ulich, Germany} \\
{\it $^{9}~$Budker Institute of Nuclear Physics, Novosibirsk, Russia} \\
{\it $^{10}$~Institute of Experimental Physics, Warsaw, Poland} \\
{\it $^{11}$~Institute of Theoretical and Experimental Physics, Moscow,
  Russia} \\ 
~~~~~~~~ \\
{\it (CELSIUS-WASA Collaboration)}}

\maketitle

{\large

\begin{center}
{\bf Abstract}\\
\medskip

All the time since its discovery the N$^*$(1440) baryon state, commonly known
as Roper 
resonance, has been a state with many question marks - despite of its 4-star
ranking in the particle data book. One reason is that it does not produce any
explicit resonance-like structures in the observables of $\pi$N or $\gamma$N
reactions. Only in partial wave analyses of $\pi$N scattering data a clear
resonance strcuture gets obvious in the $P_{11}$ partial wave.
Very recent measurements of the J/$\Psi$ decay by the BES collaboration and of
the $pp \to np\pi^+$ reaction at 1.3 GeV by the CELSIUS-WASA collaboration
show for the first time a clear resonance structure in
the invariant $n\pi^+$ mass spectrum for the Roper resonance at M $\approx$
1360 MeV with a width of about 150  
MeV. These values agree very favorably with the pole position results of
recent $\pi$N phase shift analyses. In consequence of this very low-lying pole
postion, which is roughly 100 MeV below the nominal value, the decay
properties have to be reinvestigated. From our two-pion production data we see
that the decay mainly proceeds via N$^* \to $N$\sigma$, i.e. a monopole
transition as expected for the breathing mode of the nucleon.

\end{center}

\section{Introduction} \label{s1}

Some 40 years ago, when David Roper undertook one of the first
energy-dependent phase shift analyses of $\pi$N scattering data, he discovered
\cite{roper} a resonance in the $P_{11}$ partial
wave by noting that this particular phase shift proceeds from 0$^\circ$
to 140$^\circ$ in the investigated energy range in much the same way as the
$P_{33}$ and $D_{13}$ phaseshifts do, where corresponding resonances were
already established. By looking on the energy, where the $P_{11}$ phase shift
passes through 90$^\circ$ he arrived at a mass of 1485 MeV for this new
resonance - which later on was named after him.  The peculiarity of this 
resonance, however, has been all the time, that no sign of a resonance-like
structure could  ever be observed in experimental observables, in particular
not in the total
cross section, where a resonance should show up by its Breit-Wigner-like
energy dependence. However, the Roper resonance obviously is excited in the
$\pi$N scattering process such weakly, that it is burried underneath a wealth
of other processes and can be sensed only via a very detailed partial wave
analysis - a feature, which had been noted already by Roper himself in his
paper \cite{roper}. There he had been pointing out that , whereas "previous
$\pi$N resonances were discovered from
observations on the qualitative behavior of experimental observables", the
Roper resonance "is not associated with conspicious features in the
observables measured so far". 

This situation did not change much since then, though
a huge number of very precise $\pi$N and also $\gamma$N reaction data have
been obtained meanwhile. However, what has changed is the precision of partial
wave analyses and their methods to reveal resonances. In former times usually 
a Breit-Wigner shape was fitted directly to the partial wave amplitudes
despite the fact that the latter in general still include background
terms. Such fits provide the socalled Breit-Wigner (BW) mass and width 
parameters of a resonance - see PDG \cite{pdg}. Nowadays more advanced
techniques as speed plot \cite{hoehler} and pole search techniques
\cite{man,ber} in the complex plane of the partial wave are used to deduce the
appropriate pole parameters, which should be free of background
contributions to a much higher degree and hence should represent  much more
approriately the physical 
relevant mass and width of a resonance. Whereas these two methods give similar
results for many of the resonances, they strongly 
deviate - see PDG \cite{pdg} - in case of the Roper resonance, which
in $\pi$N scattering sits upon a huge background of inelastic processes due to
pion production. This large inelasticity associated with the
Roper resonance signals already that this resonance most likely decays not
into the (elastic) $\pi$N channel but into $\pi\pi$N channels.

All recent $\pi$N partial wave analyses \cite{hoehler,cut,man,said} agree that
in case of the Roper resonance the BW method mass and width parameters are in
the range $M_{BW}$ = 1420 - 1470 MeV and $\Gamma_{BW}$ = 200 -
450 MeV, whereas the pole values are $M_{pole}$ = 1350 - 1380 MeV and
$\Gamma_{pole}$ = 160 - 220 MeV \cite{pdg} with the most recent SAID values
being $M_{pole}$ = 1357 MeV and $\Gamma_{pole}$ = 160 MeV \cite{said}. This
means that in truth the Roper resonance lies much lower than believed earlier
based on the BW method and that it is also much more narrow than believed
earlier. Its widths is now no longer exceptionally large and fits to the
typical width of nucleon excitations.

\section{Previous Attempts to  'see' the Roper Resonance}

Of course, there have been innumerable attempts to see the Roper directly in
the data, i.e. by typical signs of a resonance in the observables. Shortly
after its discovery by Roper a large number of inclusive single-arm
measurements of pion-proton and proton-proton collisions in the GeV energy
range were presented taken preferentially with magnetic spectrometers. By
varying the scattering angles the four-momentum transfer relevant for the
obtained missing mass spectra was varied accordingly. As a result such
spectra displayed a bump structure around a missing mass of 1440 MeV as long
as the four momentum transfer was very small \cite{fol,and,ede}. For larger
transfer momenta this structure disappeared. Whereas some people argued to have
visible evidence for the Roper excitation, others showed that kinemtical
reflections in sense of the Deck model \cite{deck} could be a likely
explanation for the observed bumps, too - in particular, since these
reflections should mainly appear at small momentum transfers. Indeed, it was
demonstrated by an exclusive two-pion production measurement of pp collisons
at 6.6. GeV/c that a bump between 1400 and 1500 MeV appears in the
invariant $M_{p\pi^+\pi^-}$ spectrum, if the other proton simultaneously is
excited to $\Delta^{++}$ in a peripheral collision process. I.e., the bump
occurs, if the $\pi^+$ is errornessly associated with the wrong proton - as is
the case, when constructing missing mass spectra from inclusive measurements
and thus creating the phenomenon of kinematic reflections. Hence this problem
is unavoidable in inclusive measurements.

This topic, however, came back very recently, when Morsch and Zupranski
\cite{morzup} reinvestigated the old inclusive measurements. They demonstrated
that the many of these inclusive spectra not only could 
be fitted very resonably with resonance parameters of M $\approx$ 1400 MeV and
$\Gamma \approx$ 200 MeV - i.e. parameters, which agree quite well to their
results from $\alpha$ scattering, see below - but, more essentially, that also
the strong 
momentum  transfer dependence of the observed bumps in the inclusive spectra
could be associated with the characteristics of a monopole transfer form
factor.

\section{New Generation of Experiments, which 'see' the Roper Resonance}
  
Since no possibility is known up-to-date to isolate the Roper resonance in
experimental observables of pion- and photon-induced reactions, attempts have
been  undertaken to look for its signature in other reactions, which could
filter out the 
Roper excitation in a somewhat better manner. Since the Roper resonance has
quantum numbers identical to those of the nucleon, it would be easiest to
excite it by a scalar-isoscalar probe. However, the only basic hadronic probe
of such characteristics is the $\sigma$ meson, which unfortunately decays
immediatedly into two pions, i.e. can not be used as a incident beam
particle. Hence it has been proposed to use a $\alpha$ particle beam instead
 -  so to speak as a next best choice, which is handable
 experimentally. Indeed, Morsch
et al. \cite{mor} succeeded to see the Roper resonance in the missing mass
spectrum of inelastic $\alpha$ scattering off hydrogen sitting upon background
stemming from $\Delta$ excitation in the $\alpha$ particle. Assuming a smooth
background underneath the Roper peak they extracted a mass of 1390 MeV and a
width of 190 MeV, i.e. close to the pole parameters. However, as pointed out
by the Valencia group \cite{hir} 
the $\Delta$ excitation process in the $\alpha$ particle may interfere with the
Roper excitation process in the hit proton. Taking into account such an
interference term the observed bump in the missing mass spectrum,
unfortunately, can as well be described by the conventional BW-parameters.

Very recently the BES collaboration came up with the idea to look for N$^*$
excitations in J/$\psi$ decays. The new aspect is that since J/$\psi$ is
isoscalar, the resulting $\bar{N}N^*$ and $\bar{N^*}N$ systems also have to be
isoscalar with the consequnce that only I = 1/2 nucleon excitations are allowed
in this decay. Besides a number of known higher-lying resonances also a small,
but clear peak is observed at M = 1358(6, 16) MeV with $\Gamma$ = 179(26, 50)
MeV. Since these values agree very well with the SAID pole position parameters
for the Roper resonance they associate this peak with the latter \cite{bes}.

Last but not least there is the possibility of exciting the Roper resonance
simply by nucleon-nucleon collisions at low energies - a point, which never
has been checked properly by exclusive pion-production reactions. As we have
shown \cite{wb,jp} in corresponding $\pi\pi$-production experiments close to
threshold, the inelastic collision process is governed by $\sigma$ exchange,
i.e. the Roper excitation this way may take place with virtual $\sigma$
particles as an ideal excitation probe. Besides it turns out that even in case
of pion exchange the Roper exctitation is strongly favored in the $pp
\rightarrow np\pi^+$ channel by isospin couplings. Whereas $\Delta^{++}$
excitation is expected to be the dominant structure seen in the $M_{p\pi^+}$ spectrum, the
Roper exciation should be the leading structure in the $M_{n\pi^+}$ spectrum,
since there the $\Delta^+$ excitation is suppressed by an order of magnitude.
In addition, if the reaction is carried out at sufficiently low energies, such
that only the lowest-lying $\Delta$ and $N^*$ resonances can be excited, no
kinematic reflections from higher lying resonances can contribute. Also the
formfactors of $\Delta$ and Roper should still be large enough, in order not
to suppress these excitations.

\section{Exclusive Measurements at CELSIUS-WASA}

In order to shed more light on this issue exclusive
measurements of the reactions $pp \rightarrow N\pi$ and $pp \rightarrow
NN\pi\pi$ have been carried out at several energies from 650 - 1450 MeV at the 
CELSIUS storage ring using the 
4$\pi$ WASA detector setup \cite{janucz} including the pellet target system,
see Fig.1.

\begin{figure} [t]
\begin{center}
\includegraphics[width=0.70\textwidth]{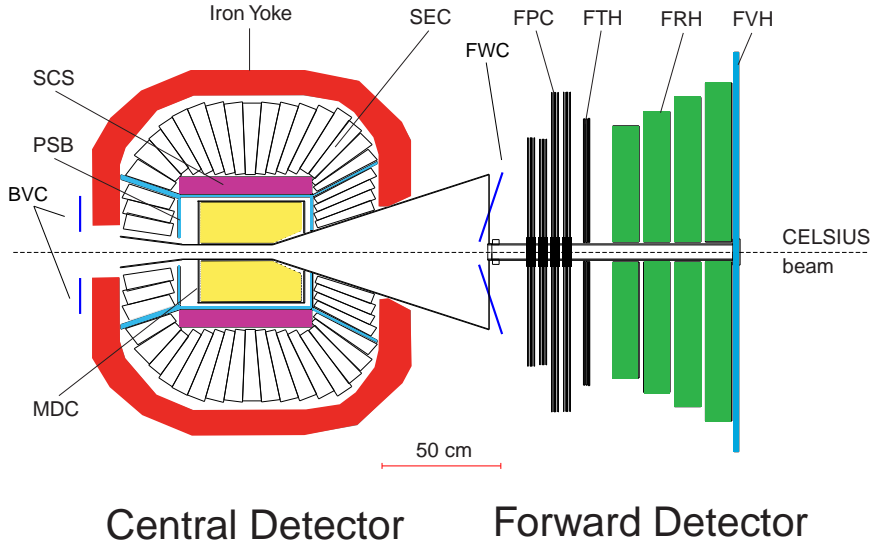} 

\caption{Side view of the WASA detector: The SuperConducting Solenoid (SCS)
  and the iron yoke for the return path of magnetic flux is shown
  shaded. Plastic scintillators are situated in the Plastic Scintillator
  Barrel (PSB), Forward Window Counters (FWC), Forward Trigger Hodoscope
  (FTH), Forward Range Hodoscope (FRH), Forward Range Intermediate Hodoscope
  (FRI), Forward Veto Hodoscope (FVH) and Backward Veto Counters (BVC). Cesium
  Iodide scintillators are situated in the Scintillator Electromagnetic
  Calorimeter (SEC). Proportional wire drift tubes, straws, make up the Mini
  Drift Chamber (MDC) and the Forward Proportional Chambers (FPC).} 

\label{fig1}
\end{center}
\end{figure}

For the reactions under consideration forward going protons have been
detected in the forward 
detector and identified by the $\Delta$E-E technique using corresponding
informations from quirl and range hodoscope, respectively.  
Charged pions, protons as well as gammas (from $\pi^0$ decay) have been
detected in the 
central detector. This way the full four-momenta have been measured for all
charged and $\pi^0$ 
particles of an event allowing thus kinematic fits with overconstraints. In
addition the direction of neutrons could be measured in most cases by their
hit pattern in forward and central detectors.

The $np\pi^+$ channel has been analyzed so far at $T_p$ = 1100 and 1300
MeV. Whereas the lower energy just suffices to reach the Roper excitation
up to its pole position, the higher energy allows to see the Roper excitation
beyond its pole. Fig. 2 shows as an example the $M_{p\pi^+}$ and 
$M_{n\pi^+}$ spectra taken at 1300 MeV. In the first one, which is purely I
=3/2, the $\Delta^{++}$ resonance is the striking feature with no other
significant signals of further resonances. A simple BW ansatz for the
$\Delta^{++}$ resonance gives a nearly perfect description for this spectrum
without any need for any additional background terms. For the  $M_{n\pi^+}$
spectrum this ansatz predicts a distribution close to phase
space. Experimentally we observe, however, a large resonance-like structure
peaking near 1350 MeV, which we associate with the Roper excitation. The
dashed lines in Fig.2 show a calculation assuming BW shapes for both the
$\Delta^{++}$ and the Roper excitation, the latter with M = 1350 MeV and
$\Gamma$ = 140 MeV. No backgound is needed for the description of the data,
which means that the obtained resonance values for the Roper may be associated
with its pole parameters.

We see that the obtained values are in excellent agreement with SAID and BES
pole parameters for the Roper resonance. For the first time such a clear Roper
excitation has been observed, its signal in the $M_{n\pi^+}$ spectrum appears
to be qualitatively as strong as that of the $\Delta^{++}$ in the
$M_{p\pi^+}$ spectrum.

\begin{figure}
\begin{center}
\includegraphics[width=18pc]{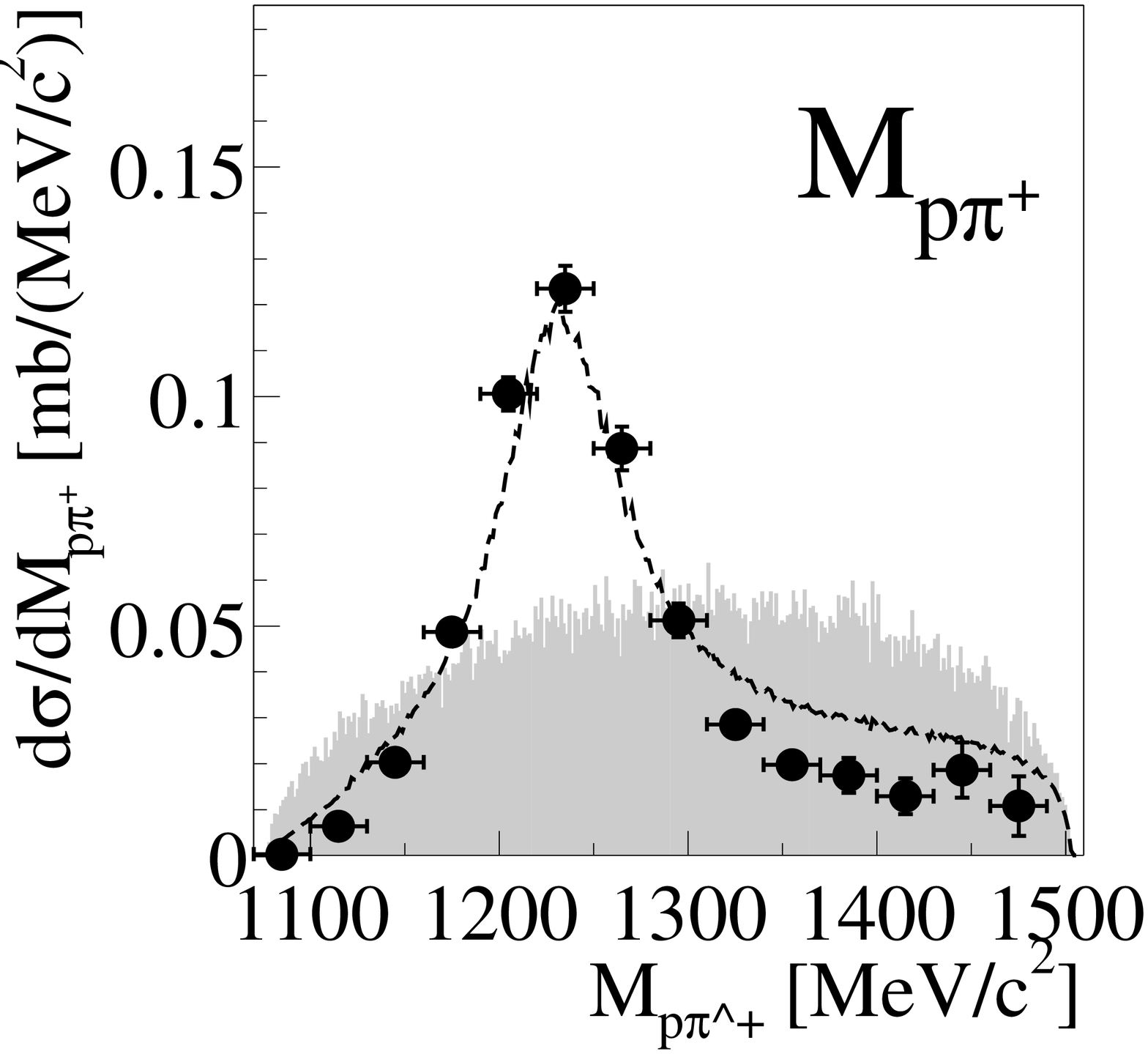}
\includegraphics[width=18pc]{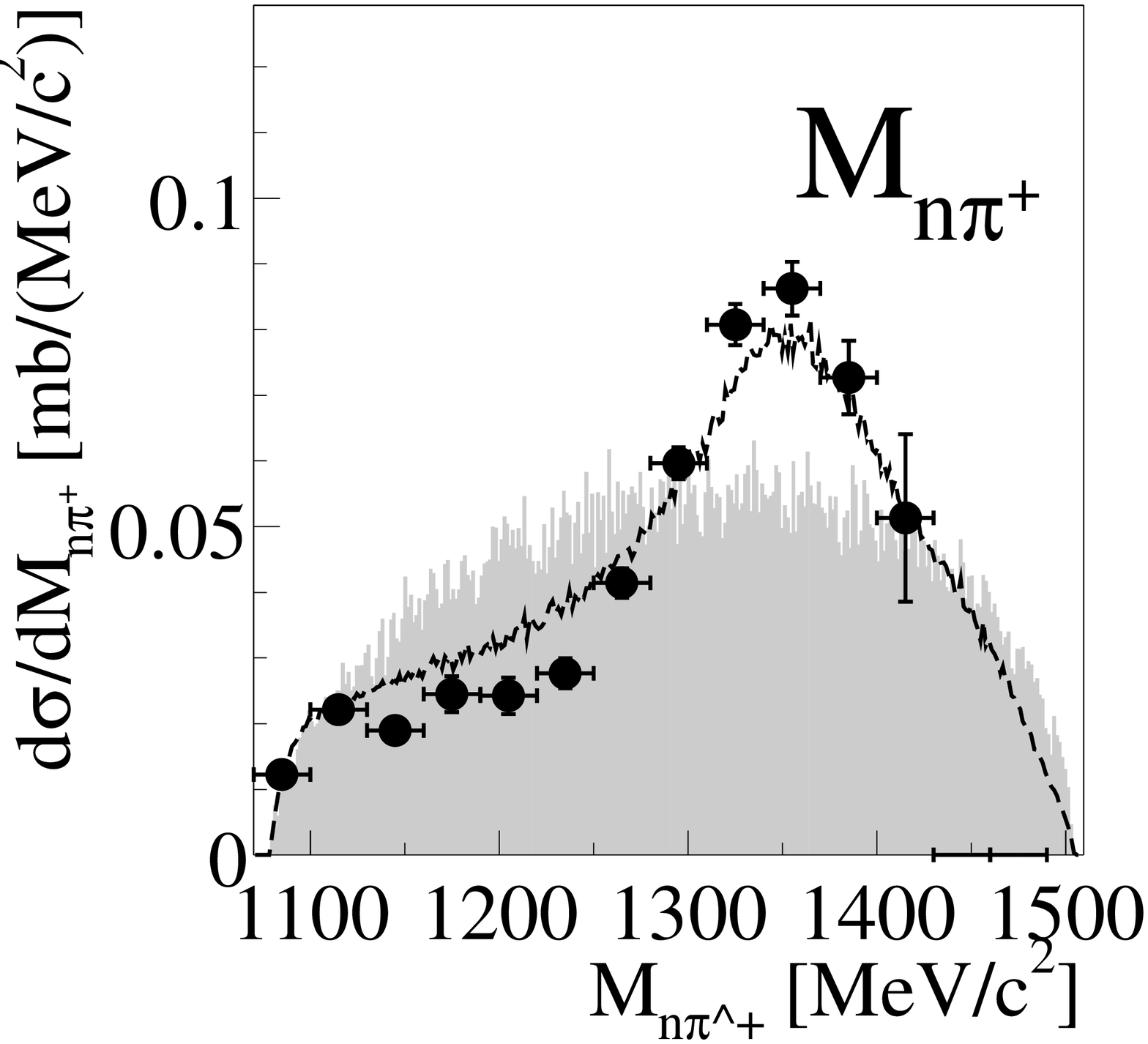}
\end{center}
\caption{Invariant mass spectra $M_{p\pi^+}$ and $M_{n\pi^+}$ obtained from
  the measurement of the $pp \to np\pi^+$ reaction at $T_p$ = 1300 MeV. The
  shaded areas show the pure phase space distributions, whereas the dashed
  lines show calculations assuming BW shapes for both the $\Delta^{++}$ and
  the Roper excitation \cite{ts}.
  }    
\end{figure}

\section{Decay Properties of the Roper Resonance}

The decay properties of the Roper resonance as given in PDG are not only very
vage, they also have been 
deduced not for the pole position of the Roper resonance, but rather for
its BW mass \cite{said,man}. The latter - as already discussed above - is much
larger than the pole value, which represents the proper resonance
mass and hence is relevant for the quotation of branching ratios, which are  
defined at $\sqrt S$ = $M_{pole}$. Since
the now established  pole mass of $M_{pole} \approx$ 1350 MeV is as much
as 100 MeV below the BW mass, this has enormeous consequences for the
branching ratios. By lowering the mass of a resonance the phase space of the
decay channels gets reduced accordingly. Such a reduction is especially severe
for decay channels, which involve finite angular momenta between the decay
products, since in such cases the decay width depends on high powers of the
available decay momenta of the emitted particles. In case of the Roper decays
this concerns in particular the decay  $N^* \to N\pi$, which procceds with a
p-wave between nucleon and pion, and still much more severely the decay $N^*
\to 
\Delta \pi \to N\pi\pi$, which proceeds with double p-wave between the two
pions and the nucleon. In addition, the new values for the pole mass of the
Roper just coincide with the 
$\Delta \pi$ branch cut, i.e. with the sum of $\Delta$ and pion masses. This
means that aside from the effect of the finite width of the $\Delta$ there is
essentially no phase space left any longer for the Roper decay into this
channel.

\begin{figure}[t]
\begin{center}
\includegraphics[width=24pc]{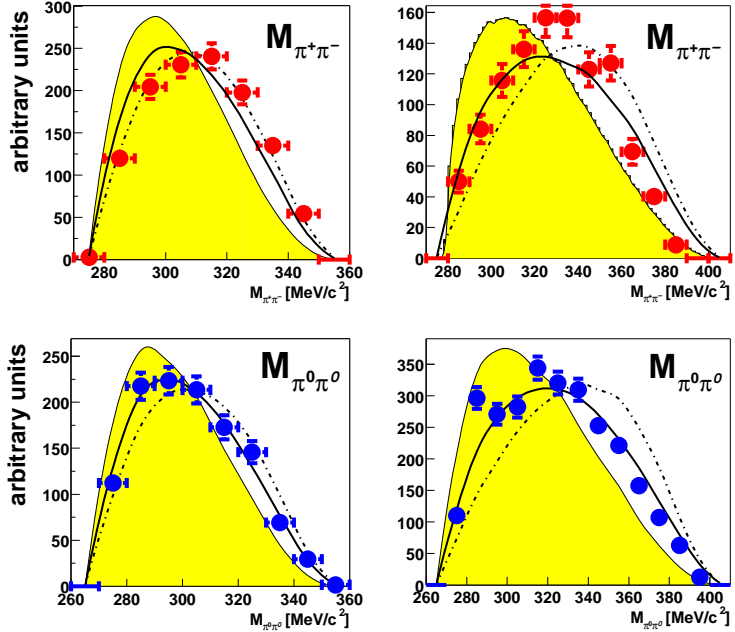}

\caption{Preliminary results for $M_{\pi^+\pi^-}$ spectra (top) and
$M_{\pi^0\pi^0}$ spectra (bottom) of the reactions $pp \to pp\pi^+\pi^-$ and
$pp \to pp\pi^0\pi^0$ at $T_p$ = 775 MeV (left) and $T_p$ = 900 MeV (right)
measured 
at CELSIUS-WASA. The data are compared to pure phase space (shaded areas) as
well as to calculations assuming either a ratio of 2:1 (dotted lines) or 8:1
(solid lines) for the branching of the Roper resonance decay into $N^* \to
N\sigma$ or alternatively into $N^* \to \Delta \pi$. Note that the current PDG value for this
branching is 1:4. From \cite{ts}.
}
\label{fig4}
\end{center}
\end{figure}

The branching ratio for the single-pion decay of the Roper can be derived
directly from the energy dependence of the imaginary part of the $P_{11}$
partial wave amplitude \cite{murat}. Using $M_{pole}$ = 1350 MeV we arrive at
about 0.3 for this branching value (compared to 0.55 - 0.75 given in PDG),
i.e. this decay can no longer be considered as the main decay branch of the
Roper decay. 

The decay of the Roper resonance into nucleon and two pions can be studied
best by two-pion production in NN-collisions. There the Roper excitation
constitutes the lowest resonance, which can contribute to this reaction. As a
consequence the near-threshoild region of this reaction is ideally suited for
the investigation of the Roper decay into two pions and nucleon
\cite{luis,wb,jp}. In addition, both decays $N^* \to N\sigma \to N\pi\pi$ and
$N^* \to \Delta \pi \to  N\pi\pi$ not only contribute to the cross section of
two-pion production, they even contribute interfering in the $\pi^+\pi^-$ and
$\pi^0\pi^0$ channels, since there they can end up in identical final states.
From the study of their interference patterns in the differential cross
sections, in particular in the $M_{\pi\pi}$ spectra, we have deduced the
relative ratio of both decay branches in exclusive near-to-threshold
measurements at $T_p$ = 750 and 775 MeV \cite{jp}. At the BW- mass this ratio
gives a value of 3.4(3) for the branching of the decay $N^* \to \Delta \pi \to
N\pi\pi$ relative to  $N^* \to N\sigma \to N\pi\pi$. This value is in
accordance with the PDG value of 4(2), though much more precise. However, if
we look at this ratio not at the BW-mass, but on the much more appropriate
pole mass of the Roper, then we arrive at values of 1.0(1), if we use
$M_{pole}$ 
= 1372 MeV - an earlier value - or 0.6(1), if we use our present value of
$M_{pole}$ = 1350 MeV. These numbers just illustrate the huge dependence of
this ratio on the pole mass. This huge dependence is - as discussed above -
due to the change in the branching of the decay via the $\Delta$ resonance,
both due to phase space shrinkage and in particular due to the involvement of
double p-waves. The results also show that at the proper Roper mass the
dominant two-pion decay channel is no longer the one via the $\Delta$, but
rather the one via the $\sigma$ channel.

The $pp\pi^+\pi^-$ channel, where the above values for the relative ratio have
been obtained, may contain both isoscalar and isovector $\pi\pi$
contributions. In order to isolate the isoscalar part we have to go to
the $\pi^0\pi^0$ channel, which is free of any isovector contributions due to
Bose symmetry. Fig. 3 shows our (preliminary) results for this channel at $T_p$
= 775 Mev and at $T_p$ = 900 MeV. Note that the Roper resonance already gets
excited kinematically up to its pole mass at $T_p$ = 900 MeV. The
shift of the data in the $M_{\pi\pi}$ spectrum relative to the distribution
for pure phase space 
originates from the interference between $N^* \to N\sigma \to N\pi\pi$ and
$N^* \to \Delta \pi \to  N\pi\pi$ routes as demonstrated in
Refs. \cite{wb,jp}. We see that in the $\pi^0\pi^0$ channel the shift and
hence the contribution from the decay via the $\Delta$ resonance
is much smaller (solid lines in Fig. 3) than expected from the analysis of the
$\pi^+\pi^-$ data (dashed lines in Fig. 3). As a
consequence  we obtain from the analysis of $\pi^0\pi^0$
channel data a value for the relative branching as low as 0.1.

From this we arrive at the following results for the branching ratios of the
Roper decay: Since the single-pion decay gives only a branching of about 0.3, 
the branching into both two-pion decay channels must be roughly 0.7. Since
furtheron the branching between $N^* \to \Delta \pi \to  N\pi\pi$ and  $N^* \to
N\sigma \to N\pi\pi$ channels is only 0.1 for the isoscalar part, this then
means that the $N\sigma$ 
decay channel is the by far largest one of all
Roper decays. From this result we see that the Roper resonance with its pole
at 1350 MeV indeed constitutes the breathing mode of the nucleon.

\section{Acknowledgements}

We acknowledge valuable discussions with H.-P. Morsch, A. Kaidalov,
V. Pugatsch and L. Dakhno on this issue.
This work has been supported by BMBF (06TU201, 06TU261), COSY-FFE, 
DFG (Europ. Graduiertenkolleg 683), Landesforschungsschwerpunkt
Baden-W\"urttemberg and the Swedish Research Council. We also acknowledge the
support from 
the European Community-Research Infrastructure Activity under FP6
"Structuring the European Research Area" programme (Hadron Physics, contract
number RII3-CT-2004-506078).

\end{document}

}

\end{document}